\documentclass[doublecol]{epl2} 

\title{A Poisson-Boltzmann Description for the Double-Layer Capacitance of an Electolytic Cell}
\shorttitle{Double-layer capacitance of an electrolytic cell}

\author{R. R. Ribeiro de Almeida\inst{1} \and L. R. Evangelista\inst{1} \and G. Barbero\inst{2}}
\shortauthor{R. R. Ribeiro de Almeida   \etal}

\institute{                    
  \inst{1} Departamento de F\'{\i}sica, Universidade
Estadual de Maring\'a -
Avenida Colombo, 5790, 87020-900,
Maring\'a, Paran\'a, Brazil.

\inst{2} Dipartimento di Fisica del
Politecnico - Corso Duca degli Abruzzi, 24,
10129 Torino, Italy.
}

\abstract{
 A Poisson-Boltzmann approach is used to determine the double-layer integral and differential capacitances in a finite-length situation for an electrolytic cell. By means of simple analytical calculations, it is shown how these quantities exhibit the bell-like and the camel-like shapes as a function of the applied voltage, when the thickness of the sample or the surface charge are varied. The problem is formulated treating the bulk liquid as a member of a grand canonical ensemble in contact with blocking electrodes. As a consequence, an applied voltage dependent  Debye's screening length arises as a fundamental length governing the electrical behavior of the system. }

\begin{document}

\maketitle

The capacitance curves in ionic liquids have typically one (bell-like shape) or two (camel-like shape) maxima,  when plotted as a function of the applied voltage.
In recent papers, this behavior has been explained as resulting from an interplay between the excluded volume effect and the effective electric constant~\cite{Lauw} or  as being essentially governed by a single parameter that represents the ratio of the bulk density of ions in the liquid to the maximal density in the double layer~\cite{Kornyshev}. In a recent article, Hatlo et al.~\cite{Lue}  considered the excess ion polarizability into the Poisson-Boltzmann theory  in order to show that the decrease in differential capacitance with voltage can be understood in terms of thickening of the double layer due to ion induced polarizability holes in water.
 In~\cite{Fawcett} was proposed a theory for concentrated electrolyte solutions in the context of lattice model and Fermi statistics to consider ions size effect for the diffuse double layer. It seems that all these important approaches invoke, in some degree, the necessity to go beyond the Poisson-Boltzmann framework to face the very rich behavior of the  experimental data regarding the double-layer capacitance in a typical electrolytic cell.  In addition to what is affirmed in these papers, here we show that by varying the thickness of the sample, some essential features of these results,
 like the camel- and bell-shaped behavior of the capacitance of an electrolytic cell as a function of the applied voltage, are obtained. The model permits to show,  in a simple manner,  how the capacitance (integral or differential)  curve of a finite sample of an insulating medium containing mobile ions can be easily determined in the framework of the Poisson-Boltzmann theory even for point-like particles. Moreover, we show also how the thickness of the sample is responsible for the shift in the maxima of the capacitance that can be found at low voltages. Recently, in Ref.~\cite{Lacoste} the  nonlinear Poisson-Boltzmann theory was invoked to investigate the behavior of a non-conductive quasi-planar lipid membrane in an electrolyte and in a static (dc) electric field.

We consider an insulating medium containing ions, with uniform dielectric coefficient $\epsilon$, in the shape of a slab, with two uniform flat surfaces in the $x-y$ plane,
separated by a distance $d$, located at $z=\pm d/2$
where blocking electrodes are placed. It is supposed, for simplicity, that all the physical quantities
entering in the calculation are only $z-$dependent. The fundamental equations have been presented in~\cite{Olivero,Libro2} but they will be summarized here for clarity. In the absence of external field the medium
is global and locally neutral such that the densities of positive and negative ions are equal to the initial
density, i.e. $n_{\pm}(z) = n_0$. In the presence of a difference of potential $U$ (supplied by an external agent)
such that $V(\pm d/2) = \pm U/2$
the equilibrium distribution of charges is governed by classical statistics, i.e.,
$ n_{\pm}(z) = n\, e^{\mp \psi(z)}$,
in which $\psi(z) = q V(z)/k_BT$ is the electric potential in $
k_BT/q $ units, where $k_B$ is the Boltzmann constant, $T$ is the absolute temperature, and $q$ is the magnitude of the elementary charge. Notice that, in the present formalism $n$ is the density of ions in the liquid phase,
to be determined by imposing the conservation of the
number of ions. This approach is equivalent to working with a fluctuating number of particles in the bulk and an unknown chemical potential that is formally determined by the conservation of the number of particles. It is just this condition that is responsible for the incorporation of a voltage dependence to the Debye's screening length, as shown below.
The spatial dependence of the  electrical potential is governed
by the Poisson's equation, which, after recognizing that the density of electric charge is given by $\rho(z) = q [n_+(z)-n_-(z)]$,   can be put in the form

\begin{equation}
\label{Poisson2}
\frac{d^2\, \psi}{dz^2} = \frac{1}{\lambda^2} \sinh\psi(z),
\quad {\rm where} \quad
\lambda^2 = \frac{n_0}{n}\, \lambda_{\rm D}^2
\end{equation}
is a length depending on the applied voltage and connected with
the Debye's screening length $\lambda_{\rm D}^2 = \epsilon k_B T/(2 q^2 n_0)$.
The nonlinear eq.~(\ref{Poisson2}) has to be solved with the boundary conditions
$\psi(\pm d/2) = \pm u = q U/ (2 k_B T)$,
and can be integrated to give

\begin{equation}
\label{integral}
\frac{d\psi}{dz} = \frac{\sqrt{2}}{\lambda}\,\sqrt{\cosh\psi + k},
\end{equation}
where $k$ is an integration constant.
The conservation of the number of particles is given by
\begin{equation}
\label{conserva}
n_0\, d = \int_{-d/2}^{d/2} n_+(z)\, dz = \frac{\lambda}{\sqrt{2}} \, n\, \int_{-u}^{u}
\frac{e^{-\psi}}{\sqrt{\cosh\psi + k}}\, d\psi
\end{equation}
and can be rewritten as

\begin{equation}
\label{jota}
\int_{-u}^u \frac{e^{-\psi}}{\sqrt{\cosh\psi + k}}\, d\psi = J(k,u) = \sqrt{2} \frac{\lambda\, d}{\lambda_{\rm D}^2},
\end{equation}
whereas eq.~(\ref{integral}) becomes

\begin{equation}
\label{isao}
\int_{-u}^u \frac{1}{\sqrt{\cosh\psi + k}}\, d\psi = I(k,u) = \sqrt{2} \frac{d}{\lambda},
\end{equation}
which, together, imply that

\begin{equation}
\label{k}
I(k,u) J(k,u) = 2 \left(\frac{d}{\lambda_{\rm D}}\right)^2.
\end{equation}

The
fundamental equations have been established and the procedure is now straightforward.
Indeed, $I(k,u)$ and $J(k,u)$ can be expressed in terms of elliptic integrals of first and second
kind and, for a given applied voltage $U$, from eq.~(\ref{k}), one obtains
$k(U)$ for a sample of definite thickness $d$ and a Debye's screening length $\lambda_{\rm D}$.
Once this is done, the effective (voltage dependent) Debye screening length can determined by the expression~(\ref{isao}).
Finally, with these quantities determined as a function of $U$, the electric field profile is then easily
obtained from eq.~(\ref{integral}), i.e., $E(z) = - (k_B T/q )\, d\psi/dz$.

\begin{figure}
\includegraphics*[scale=.1,angle=0]{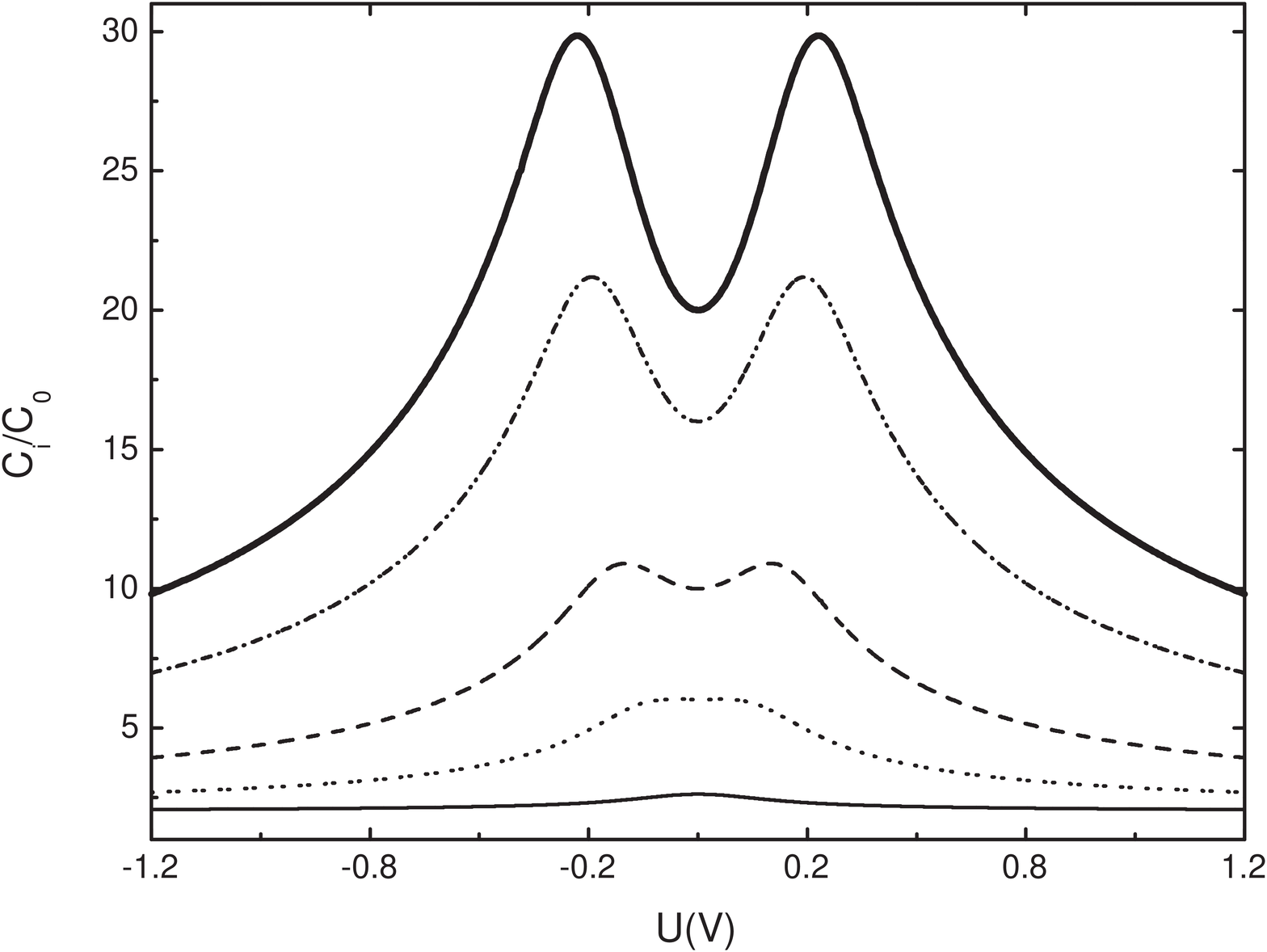}
\caption{Integral double-layer capacitance vs. the bias voltage, $U$,  for different thicknesses of the sample: $d = 10 \, \mu$m (thick solid), $d = 8 \, \mu$m (dashed-dotted), $d = 5 \, \mu$m (dashed),
$d = 3 \, \mu$m (dotted), and
$d=1\, \mu$m (thin solid). The curves have been drawn for $\lambda_{\rm D} =0.5\,\mu$m. }
\label{Fig_Cap_Integral}
\end{figure}

\begin{figure}
\includegraphics*[scale=.08,angle=0]{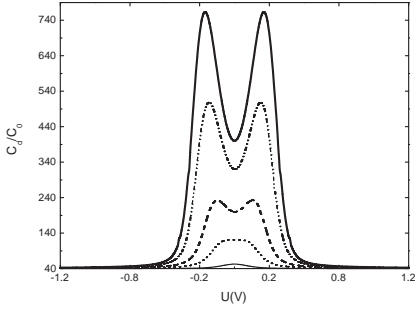}
\caption{Differential double-layer capacitance vs. the bias voltage, $U$,  for different thicknesses of the sample according to the convention adopted in fig.~\ref{Fig_Cap_Integral}. }
\label{Fig_Cap_Differential}
\end{figure}

\begin{figure}
\includegraphics*[scale=.08,angle=0]{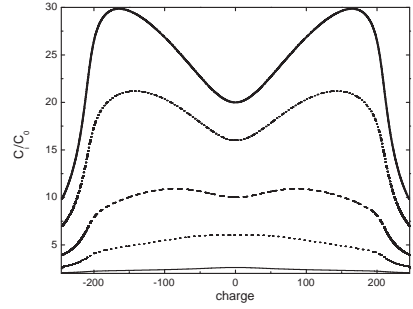}
\caption{Integral double-layer capacitance vs. surface charge,  for different thicknesses of the sample as in figs.~\ref{Fig_Cap_Integral} and~\ref{Fig_Cap_Differential}. }
\label{Fig_Surface_Charge}
\end{figure}

To obtain simple expressions for the double-layer capacitance, it is necessary to calculate the net charge on the surfaces, given by
$\sigma = \epsilon E (z=-d/2)$. In this manner, the differential double-layer capacitance is given by

\begin{equation}
\label{c_differential}
C_d = \frac{d\sigma}{dU} = C_0 \frac{d}{du} \left[I(k, u) \sqrt{\cosh u + k}\right],
\end{equation}
in which $C_0= \epsilon/ 2 d$ and~(\ref{integral}) and (\ref{isao}) have been used. In the same manner, the integral double-layer capacitance will be given by

\begin{equation}
\label{c_integral}
C_i = \frac{\sigma}{U} = C_0 \frac{I(k, u) }{u}\, \sqrt{\cosh u + k}.
\end{equation}
In fig.~(\ref{Fig_Cap_Integral}), the integral capacitance is shown as a function of the external voltage. The camel-like shape is evident, but it is strongly dependent on the ratio between the thickness of the sample and the Debye length $\lambda_{\rm D}$. Indeed, when $ d\simeq \lambda_{\rm D}$, the maximum is found for $U=0$ (see, e.g., the thin solid line). The camel-like shape is found if $d > \lambda_{\rm D}$.
Likewise, the differential double-layer capacitance, shown in fig.~\ref{Fig_Cap_Differential}, exhibits similar behavior with the thickness. In fig.~\ref{Fig_Surface_Charge}, the integral double-layer capacitance is shown as a function of the charge $\sigma$  for different thicknesses of the sample.
The analysis presented above is based on the conservation of the number of
particles in the capacitor, eq.~(3). This means that, at first sight,  we are considering a closed system, where the number of ions is fixed by its initial volume, as in a sealed condenser. In this manner, when the system is subject to
higher voltages the amount of net charge accumulated in each double
layer will eventually stop growing with further increase of voltage,
and the capacitance will decrease. For this reasons
we observe that in larger systems the falling branches of
capacitance emerge at larger voltages. The application of our model to real experiments, where the electrodes are embedded into a sea of electrolyte, and can suck ions from the bulk, requires some caution. In fact, in this case the system is no longer closed, and it can exchange particles with the reservoir. However, we are working here in the slab approximation, i.e., the electrolyte is limited by two flat, infinitely large, electrodes. This means that our sample is isolated, as the  semi-infinite samples considered in various recent works~\cite{Kornyshev,Henderson,Lamperski}. It follows that our theoretical predictions can be compared with the ones reported in~\cite{Kornyshev,Henderson,Lamperski}.  Since the quoted theoretical model~\cite{Kornyshev,Henderson,Lamperski} have been used to interpret experimental data relevant to an open condenser, where the electrodes can adsorb ions coming from outside of the sample, we can conclude as follows. In real case, the systems are not isolated, but the dominant role on the electrical response of the electrolytic cell is played by the ions contained in the volume limited by the electrodes. In the opposite case, the theoretical description proposed by us, as the one following from the models discussed in~\cite{Kornyshev,Henderson,Lamperski}, does not work, and the description has to include also the lateral ionic current. However, at the present, due to the good agreement between theoretical predictions and experimental data it seems that the effect related to these horizontal current can be neglected.

\begin{figure}
\includegraphics*[scale=.08,angle=0]{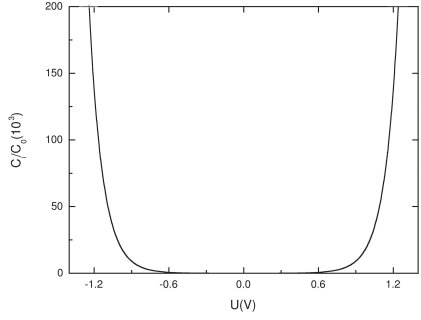}
\caption{Integral double-layer capacitance vs. the applied voltage,  for  $d= 10\,\mu$m and $\lambda_{\rm D}= 0.5\, \mu$m, when $ \lambda= \lambda_{\rm D}$. }
\label{Fig_Cap_Gouy}
\end{figure}

The behaviors shown in fig.~1, fig.~2 and in fig.~3 are consistent with the results found in~\cite{Henderson} as a function of the concentration and with the Monte Carlo simulation in~\cite{Lamperski}. The simulation were performed in the grand canonical ensemble to remove uncertainties around the determination of the bulk concentration.  Likewise, in our system, for a fixed value of $n_0$, the actual bulk concentration depends on $n$ which, in turn, is fixed by imposing the conservation of the number of particles. This means that our approach treats the liquid system as a member of a grand canonical ensemble as well, i.e., as a system having a variable number of particles and, from the point of view of the ensemble theory, it is equivalent to have as macroscopic variables $T, V, \mu$, where $\mu$ is the chemical potential.
The complete system, i.e., the liquid medium plus the walls (solid phase) is, instead, characterized by the equilibrium variables $T,V,N$, i.e., it can be treated in the framework of the canonical ensemble.
In addition, we notice that a bell-like shape curve similar to the one shown in figs.~(\ref{Fig_Cap_Integral}) and (\ref{Fig_Cap_Differential}) has been found in the framework of a mean-field approach at higher concentration~\cite{Fedorov2} and by means a Landau-Ginzburg-type continuum theory~\cite{Bazant}, which takes into account the exclusion volume corrections. At low concentrations, a camel-like shape capacitance curve was reported in~\cite{Kornyshev,Lauw}.
In our approach, illustraded here for parameters more appropriated to ordinary electrolytes, these behaviors are found when
the ratio $d/\lambda_D$ varies. In fact, when $d/\lambda_D >1$ (which corresponds to a more diluted system), we obtain the camel-shape behavior. For $d \simeq \lambda_D$, the model predicts a bell-shape behavior.
Finally, in fig.~\ref{Fig_Cap_Gouy}, the integral double-layer capacitance as a function of the external voltage is shown in the limiting situation for which $\lambda \to \lambda_{\rm D}$, and  exhibits the expected classical Gouy-Chapman behavior~\cite{Gouy,Champman,Bockris,Schmickler}.
In this case, the voltage dependence introduced in $\lambda(U)$ by means of the condition of conservation of number of particles is lost,  as can be deduced  by comparing the behaviors shown in figs.~\ref{Fig_Cap_Integral} and ~\ref{Fig_Cap_Gouy}. In fig.~\ref{Fig_Cap_Integral}, the effective Debye's length depends on $U$ and we can found both, the bell- and camel-shape behavior as $d/\lambda_D$ increases. In fig.~\ref{Fig_Cap_Gouy},
$\lambda=\lambda_D$, and the curve presents the Gouy-Chapman behavior, i.e., it is minimum at the point of zero charge.

In the high voltage limit, i.e., $u \gg 1$, it is possible to show that the electric potential across the sample becomes~\cite{Olivero}:

$$
\psi(z) = 2 u \frac{z}{d} \left[1- \frac{1}{u^2} D^2 \right] + 2 \frac{e^{-u}}{u } D^2\, \sinh\left(\frac{2 u z}{d}\right),
$$
where  $D = d/(2 \lambda_{\rm D}) $. The differential double-layer capacitance is now given by

$$
C_d = 2 C_0 \left[1+ D^2\left(\frac{1}{u^2} - 2 e^{- 2u}\right)\right],
$$
whereas the integral capacitance becomes

$$
C_i = 2 C_0 \left[1+ \frac{D^2}{u} \left( 1- \frac{1}{u} + e^{- 2u}\right)\right].
$$
In the limiting situation of small applied voltage, i.e., $u \ll 1$,  the electric potential
may be approximated as~\cite{Olivero}:

$$
\psi(z) \approx u \frac{\sinh(z/\lambda_{\rm D})}{\sinh(d/2\lambda_{\rm D})},
$$
and the net charge on the surface is given by $\sigma = (\epsilon/ 2\lambda_{\rm D}) U \coth D$,   and, in this case, both the differential and the integral capacitance coincides, namely

$$
C_d = C_i = C_{\rm GC} \coth D \approx C_{\rm GC},
$$
in which $C_{\rm GC} = (\epsilon /2 \lambda_{\rm D})$ is the linear Gouy-Chapman or ``Debye'' capacitance~\cite{Schmickler}, as expected because, typically,  $D \gg 1$, and $\coth D \approx 1$.

\begin{figure}
\includegraphics*[scale=.08,angle=0]{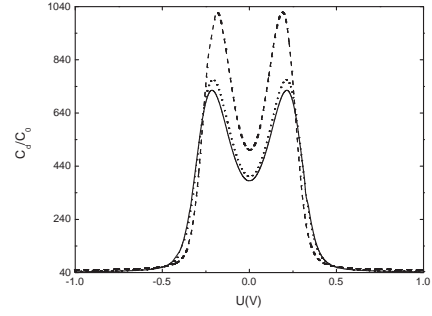}
\caption{Differential double-layer capacitance vs. the applied voltage,  for  $d= 10\,\mu$m and $\lambda_{\rm D}= 0.5\, \mu$m,  for three different temperatures: $\rm{T} = 250 \,$\rm{K} (dashed), $\rm{T} = 290 \,$\rm{K} (dotted) and $\rm{T} = 298.15 \,$\rm{K} (solid).  }
\label{Fig_Cap_Temp}
\end{figure}

Actually, in the limit of very large separation between the electrodes, the sample can be considered as formed by two half spaces, as shown in~\cite{Olivero}, and the electric potential may be written as $\psi(z) = \psi_+(z) + \psi_-(z)$, where

\begin{equation}
\psi_{\pm } = 2 \ln \left[\frac{1 \pm \gamma e^{(z \mp d/2)/\lambda_D}}{1 \mp \gamma e^{(z \mp d/2)/\lambda_D}}\right], \quad {\rm with} \quad \gamma = \tanh\left(\frac{u}{4}\right).
\end{equation}
By using eqs.~(\ref{c_differential}) and~(\ref{c_integral}), this exact limiting expression leads to

$$
C_d = \frac{1}{2} C_{GC} \cosh\left(\frac{u}{2}\right) \quad {\rm and} \quad
C_i = \frac{1}{2} C_{GC} \frac{\sinh\left(\frac{u}{2}\right)}{u},
$$
for $D \gg 1$. Thus, for very large separation between the electrodes, one concludes that the capacitances are independent of this separation. Moreover, for $u=0$,  both the differential as well as the integral capacitances reduce to the half of the Couy-Chapmann capacitance corresponding to a semi-infinite sample.  This is not surprising, because for very large sample one should not expect to observe a difference between the grand-canonical and canonical ensembles, since most of the ions are coming from the bulk. To complete the analysis, in fig.~(\ref{Fig_Cap_Temp}), the differential double-layer capacitance is shown as a function of the applied voltage for different values of the temperature. It diminishes with increasing temperature as found in other approaches~\cite{Lamperski, Klos}.

 Even if it is true that electrochemical process, such as reactions, are determined by the ion profiles~\cite{Henderson2}, and the capacitance is some kind of integral of ion profiles, its determination is crucial for investigating the response of the
electrolytic cell to an external applied voltage. The very rich behavior of the capacitance is a confirmation of its important role in the electrical behavior of the cell. In this sense, the efforts concentrated in the understanding of this complex behavior are justified. Here, we have presented a simple analysis giving a complete profile for the behavior of the double-layer capacitance in a finite-length situation in the framework of the Poisson-Boltzmann statistics. For explaining the transition from the bell-shape to the camel-shape behavior, it is crucial to describe in the correct manner how the bulk concentration of particles are varying in the presence of the applied voltage. In the formalism we have presented, this is done by considering a fluctuating bulk density which is fixed by the imposition of the conservation of the number of particles. More precisely, even working with blocking electrodes, this formalism permits one to treat the charges accumulated at the electrodes  as pertaining to the
surface of the system, instead of remaining in the bulk. As a consequence, in the system appears a Debye's screening length that depends on the bias voltage. Only in the absence of external field, this quantity coincides with the usual Debye's screening length.
It is then possible to account for the effect of the external voltage by considering the role of two important lengths characterizing the system, namely, the thickness defining the geometry of the cell, and an  ``effective'' electric length strongly connected with the Debye's classical length. In this manner, for $d \simeq \lambda_D$, the system presents a bell-shaped (a maximum at the point of zero charge) behavior. As  $d$ increases, a shallow minimum appears at the point of zero charge and two other symmetric maxima appear in the region of low voltage (camel-shaped behavior). The shallow minimum evolves to a deep minimum when $d \gg \lambda_D$.
 As a final remark, it is worth mentioning that all the more sophisticated approaches that have been proposed in the last years to face the problem have enlarged a lot the comprehension of the problem and new efforts have to be made to push further this comprehension. However, a global understanding of the electrical behavior of the sample that takes into account the integrated ion profiles can be easily achieved by means of a simple Poisson-Boltzmann approach formulated in the framework of the grand canonical ensemble.

\acknowledgments

This work was partially supported by the National Institutes of Science and Technology of Complex Fluids -- INCT-FCx (L. R. E.).

\end{document}